\documentclass[sigconf,authorversion,nonacm]{acmart}

\makeatletter
\def\@ACM@checkaffil{
    \if@ACM@instpresent\else
    \ClassWarningNoLine{\@classname}{No institution present for an affiliation}%
    \fi
    \if@ACM@citypresent\else
    \ClassWarningNoLine{\@classname}{No city present for an affiliation}%
    \fi
    \if@ACM@countrypresent\else
        \ClassWarningNoLine{\@classname}{No country present for an affiliation}%
    \fi
}
\makeatother

\AtBeginDocument{%
  \providecommand\BibTeX{{%
    \normalfont B\kern-0.5em{\scshape i\kern-0.25em b}\kern-0.8em\TeX}}}

\usepackage[english]{babel}
\usepackage{blindtext}

\usepackage[T1]{fontenc}
\usepackage[font=small,labelfont=bf,tableposition=top]{caption}
\usepackage{longtable}

\DeclareCaptionLabelFormat{andtable}{#1~#2  \&  \tablename~\thetable}
\begin{document}


\title{Expanding the Role of Affective Phenomena\protect\\in Multimodal Interaction Research}
 
\author{Leena Mathur}
\affiliation{%
  \institution{School of Computer Science\\Carnegie Mellon University}}
\email{lmathur@cs.cmu.edu}

\author{Maja J Matarić}
\affiliation{%
  \institution{Department of Computer Science\\University of Southern California}}
\email{mataric@usc.edu}

\author{Louis-Philippe Morency}
\affiliation{%
  \institution{School of Computer Science\\Carnegie Mellon University}}
\email{morency@cs.cmu.edu}



\begin{abstract}
In recent decades, the field of affective computing has made substantial progress in advancing
the ability of AI systems to recognize and express affective phenomena, such as affect and emotions,  during human-human and human-machine interactions. This paper describes our examination of research at the intersection of multimodal interaction and affective computing, with the objective of observing trends and identifying understudied areas. We examined over 16,000 papers from selected conferences in multimodal interaction, affective computing, and natural language processing: ACM International Conference on Multimodal
Interaction, AAAC International Conference on Affective Computing and Intelligent Interaction, Annual Meeting of the Association for Computational Linguistics, and Conference on Empirical Methods in Natural Language Processing. We identified 910 affect-related papers and present our analysis of the role of affective phenomena in these papers. We find that this body of research has primarily focused on enabling machines to recognize and express affect and emotion. However, we find limited research on how affect and emotion predictions might be used by AI systems to enhance machine understanding of human social behaviors and cognitive states. Based on our analysis, we discuss directions to expand the role of affective phenomena in multimodal interaction research. 
\end{abstract}



\maketitle

\section{Introduction}
In recent decades, research in psychology and neuroscience has highlighted the importance of affective phenomena in understanding, explaining, and predicting how humans behave and think during real-world social interactions \cite{dukes2021rise, forgas2012affect}. This body of research has demonstrated explanatory relationships among affective phenomena (e.g., affect and emotion), cognitive processes (e.g,. memory, attention, perception, and decision-making), and behavioral processes (e.g., habits, adaptation, stimulus-response actions) \cite{dukes2021rise}. In addition, affective states have been shown to regulate the dynamics of human \textit{social behaviors} (e.g., communicative social signals) and \textit{cognitive states} (e.g., attitudes) during social interactions \cite{moore1990affect, isen1987positive, clore2007affective}. 

In parallel to the aforementioned progress in the affective sciences, recent decades of computer science research have laid foundations in affective computing \cite{picard2000affective, pantic2003toward, d2015review}, with substantial progress in advancing the ability of AI systems to estimate affective phenomena in humans. After affective phenomena have been predicted by an AI system, we believe those predictions can be used to enhance the system's understanding of human social behaviors and cognitive states, towards more socially-intelligent AI. We were, therefore, motivated to explore the question: \textit{How, and to what extent, have affective phenomena been used by AI systems in multimodal interaction research to enhance machine understanding of human social behaviors and cognitive states?}

To begin answering this question, we scoped a study to examine trends in how multimodal interaction research has treated the role of affect and emotion in over 16,000 papers selected from premier conferences that represent communities in multimodal interaction, affective computing, and natural language processing (NLP): ACM International Conference on Multimodal Interaction (ICMI), AAAC International Conference on Affective Computing and Intelligent Interaction (ACII), Annual Meeting of the Association for Computational Linguistics (ACL), and Conference on Empirical Methods in Natural Language Processing (EMNLP).  
 
Our paper makes four contributions. First, we identify 910 papers related to affect and emotion from past decades of proceedings at ICMI, ACII, ACL, and EMNLP and categorize the role of affect and emotion in these papers. Second, we quantify the extent to which affect and emotion have been used by AI systems to enhance AI system understanding of human social behaviors and cognitive states in these 910 papers. Third, we analyze trends in how affect and emotion have been used to enhance machine understanding of human social behaviors and cognitive states.
Fourth, based on our analysis, we offer insights into future directions to expand the role of affect and emotion in multimodal interaction research.




\section{Affective Phenomena, Social Behaviors, and Cognitive States}
\label{abc}
\begin{figure}[!t]
  \includegraphics[width=0.8\linewidth]{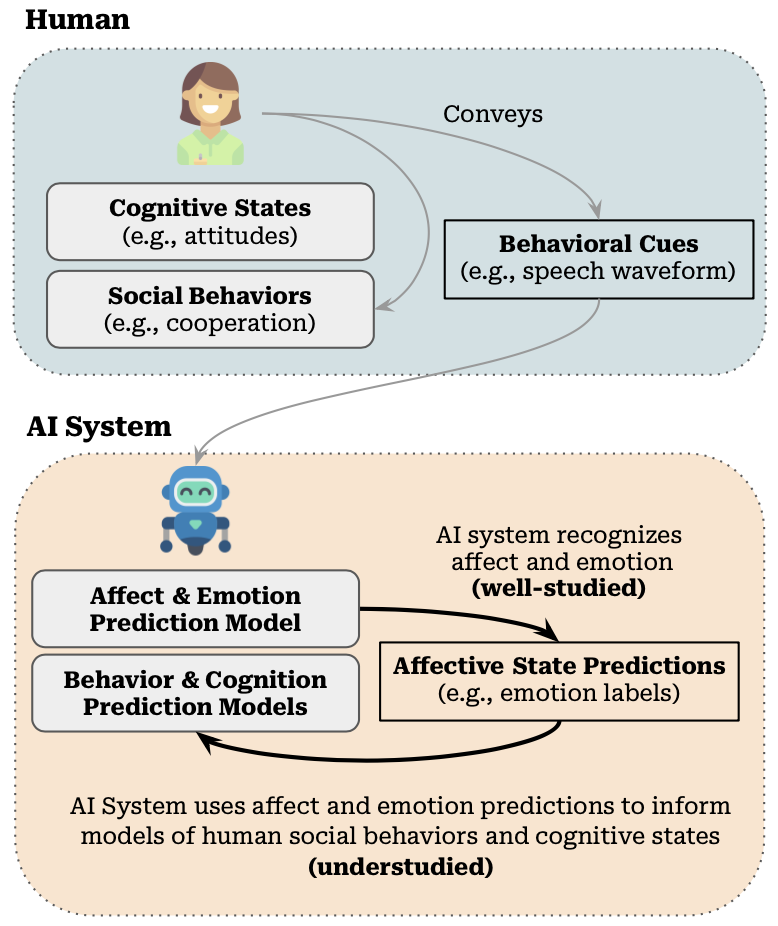}
  \caption{Conceptualization of use of an AI system to predict affect and emotion in a human and, then, use those predictions in models of human social behaviors and cognitive states.}
  \label{fig:teaser}
\end{figure}

This section provides a brief overview of relationships among affective phenomena, social behaviors, and cognitive states. This is a growing research area in psychology and neuroscience \cite{ dukes2021rise}. Findings suggest that affective phenomena (e.g., affect, emotion) drive human social behavior and social cognition  \cite{adolphs2001interaction, ito2001affect, bechara1994insensitivity, niedenthal2000emotional}. Affective processes can influence how people remember social information \cite{lemerise2000integrated}, make decisions \cite{forgas2012affect, isen1983influence}, and perceive others during interpersonal social interactions; for example, a person's affective state before an interpersonal interaction can influence whether or not they end up liking or favourably judging another person (the reinforcement-affect model) \cite{clore1974reinforcement, clore1974knowing}. Prominent models backed by empirical findings include the \textit{affect-as-information} theory (the perspective that humans directly query their affective state when making judgements) \cite{clore2001affect, clore2013affective}, the \textit{affect priming} theory (the perspective that affect primes connections across concepts during reasoning and choice of behaviors in social situations) \cite{klauer2003affective}, and the \textit{affect infusion model} (defines social contexts in which affect influences the choice of social behavior) \cite{forgas2013affect}. These  relationships among affective phenomena, social behaviors, and cognitive states can be leveraged by AI systems during real-world interactions. As conceptualized in \textbf{Figure \ref{fig:teaser}}, an AI system that uses behavioral cues to predict a human's affective state can then use those predictions to model the human's social behaviors and cognitive states.



\raggedbottom

\section{Selecting and Categorizing Papers}
\label{methods}
This study was designed to capture and analyze broad trends in  premier conference venues, selected to reflect communities in multimodal interaction (ICMI), affective computing (ACII), and NLP (ACL and EMNLP). We note that our choice to include NLP venues in our analysis is due to the dialogue, discourse, and interaction research in this community. Our inclusion of ACL also encompassed prior papers at regional ACL conferences NAACL, EACL, AACL.

Data were accessed from the available online conference proceedings: ICMI proceedings began from 2002\footnote{\url{https://dl.acm.org/conference/icmi-mlmi/proceedings}}, ACII  proceedings began from 2009\footnote{\url{https://ieeexplore.ieee.org/xpl/conhome/1002992/all-proceedings}}, and ACL/EMNLP proceedings began from 1979 in the ACL Anthology Corpus dataset\footnote{\url{https://huggingface.co/datasets/ACL-OCL/acl-anthology-corpus}}. Our search yielded \textbf{16,966 papers}: 1161 ICMI papers from 2002-2022, 786 ACII papers from 2009-2022, and 15019 ACL and EMNLP papers from 1979-2022. 

We applied an initial filter to select papers in which the title or abstract contained at least one of the following keywords:\textit{ affective, affect-aware, valence, arousal, positive affect, negative affect, emotion, emotions, emotion-aware, emotional}. We chose this approach to filter our initial set of papers on the assumption that papers disseminating findings applicable to affective phenomena during social interactions will include at least one of these terms in the title or abstract; filtering papers with this approach should effectively capture affect-related  research (papers that address affect and emotion) for our further analysis. This yielded a total of \textbf{910 papers}  (129 ICMI papers, 547 ACII papers, 234  ACL/EMNLP papers).

We examined the 910 papers and categorized them into the following 7 groups, based on the primary focus of each paper in its treatment of affect and emotion.

\textbf{(1) Recognizing Affect and Emotion}: This category includes papers on modeling efforts to predict affect and emotion. For example, a paper that proposes a method to predict valence and arousal labels for speakers in a video dataset would be in this category. 

\textbf{(2) Expressing Affect and Emotion}: This category includes papers that focus on techniques to enable virtual and embodied AI agents to express affect and emotion. For example, a paper that proposes a method to express facial emotions in virtual human avatars would be in this category.

\textbf{(3) Recognizing and Expressing Affect and Emotion}: This category includes papers that perform both recognition and expression of affect and emotion, warranting a separate category. For example, a paper that proposes a method to recognize emotional states in a human speaker and uses that method to inform a virtual avatar's expressed emotion would be in this category.
    
\textbf{(4) Using Affect and Emotion for Enhanced Machine Understanding of Human Social Behaviors and Cognitive States}: This category includes papers that explore the role of affect and emotion to enhance machine understanding of human social behaviors and cognitive states during interactions. For example, a paper that uses the outputs of an affect prediction model to predict human social behaviors would be in this category. 

\textbf{5) Affect and Emotion Frameworks and  Analysis}: This category includes conceptual work and psychology studies of humans during interactions. For example, a paper analyzes students' emotions while playing a game would be in this category.

\textbf{(6) Tools, Interfaces, and Datasets}: This category includes papers that discuss data collection tools, papers on interfaces for facilitating interactions, and papers that introduce datasets. 

\textbf{(7) Miscellaneous}: This category includes papers that did not fit into prior categories. For example, a paper that focused on techniques for video retrieval and emphasized that the video's social context involved emotion, would be in this category.

\section{Distribution of Research Focus in the Affect-Related Papers}
\label{prior_trends}
\begin{figure}[!t]
    \centering
    \includegraphics[width=0.7\linewidth]{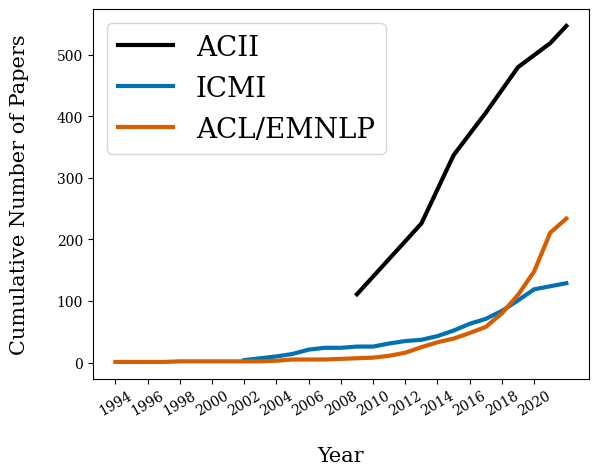}
    \caption{Accumulation of affect-related papers at ACII, ICMI, and ACL/EMNLP over time.}
    \label{fig:cumulative}
\end{figure}

All 910 affect-related papers were published between 1994-2022. The number of these papers published across time at the studied venues is visualized in \textbf{Figure \ref{fig:cumulative}}. We observe a substantial increase in the number of papers during the past decade at ICMI and ACII, and during the past 5 years at ACL/EMNLP. We note that this increased interest in studying affective phenomena complements the acceleration of research activity in affective phenomena across psychology, neuroscience, humanities, and social sciences \cite{dukes2021rise}. 

\begin{figure}[t]
    \centering
    \includegraphics[width=0.9\linewidth]{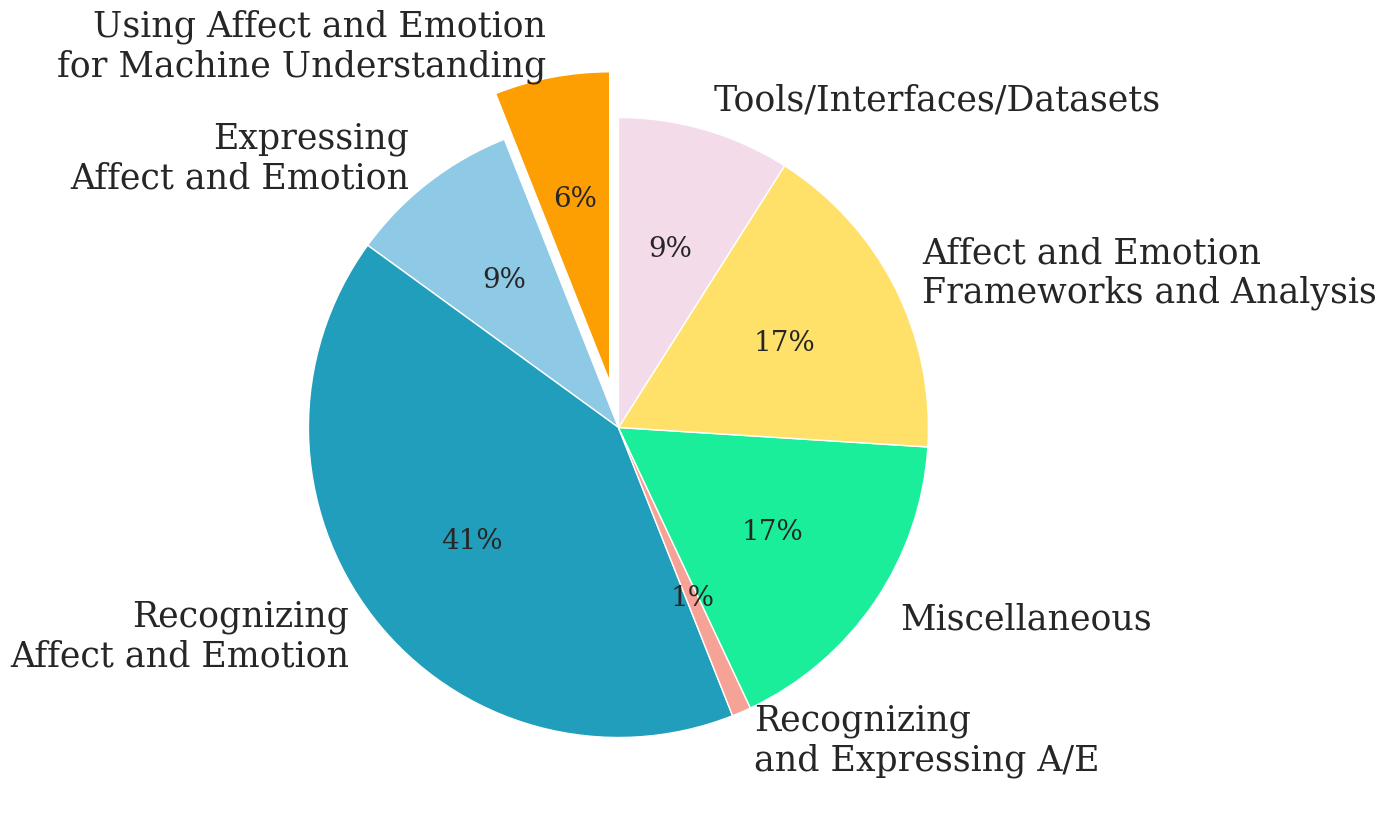}
   
    \caption{Distribution of research focus in affect-related papers.}
    \label{overall_categories}
  \end{figure}

The distribution of research focus in the 910 papers is visualized in \textbf{Figure \ref{overall_categories}}. We find that across these papers, the large proportion focused on techniques for machines to recognize and express affect and emotion (41\% and 9\%, respectively); an additional 1\% focused on both these challenges. Only 6\% (52 papers) discussed research that investigated using affect and emotion to enhance machine understanding of human social behaviors and cognitive states. We analyze this subset of 52 papers in \textbf{Section \ref{the_func_papers}} for insights about this understudied area. Among the remaining papers, 17\% focused on affect-related  frameworks and analysis, 9\% on new tools, interfaces, and datasets, and 17\% on miscellaneous topics (e.g., video retrieval in papers that happened to mention emotion).

\section{Use of Affect and Emotion for Enhanced Machine Understanding}
\label{the_func_papers}
\begin{figure}[t]
    \centering
    \includegraphics[width=0.7\linewidth]{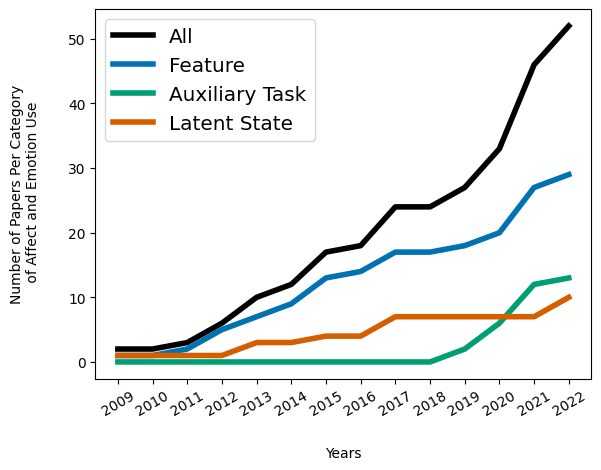}
    \caption{Accumulation of papers that used affective phenomena for enhanced machine understanding, split across use as a feature, in auxiliary tasks, and as a latent state.}
    \label{fig:cumulative_func}
\end{figure}

We analyzed the 52 papers that used affect and emotion to enhance machine understanding of social behaviors and cognitive states. We find that these papers used affective phenomena in three primary ways: as a \textit{feature}, in an \textit{auxiliary task}, and as a \textit{latent state}. All 52 papers were published between 2009-2022; 54\% of them were published in the last 4 years. The  accelerating increase in papers on this topic is visualized in \textbf{Figure \ref{fig:cumulative_func}}. This trend demonstrates a growing interest across the multimodal interaction, affective computing, and NLP communities in this understudied research area. We observed a steady increase in papers that used affective phenomena as a feature and a slower increase in papers that used affective phenomena as a latent state. In the past 5 years, we also observed a sharp increase in the number of papers that used affective phenomena in an auxiliary task. We further examine these categories. 

\textbf{Affective Phenomena as Features:} 29 of the 52 papers used affective phenomena as features to predict human social behaviors and cognitive states. In these papers, affective phenomena were used to predict the following social behaviors and social signal dynamics in human-human and human-machine interaction: head nods \cite{lee2009learning}, humor \cite{yang2021choral}, idiom and metaphor expression \cite{peng2018classifying, jang2016metaphor}, non-cooperative behavior \cite{stratou2015emotional},  deception \cite{mathur2020introducing}, fake communication \cite{chawla2021towards}, self-disclosure \cite{bak2012self}, intimacy \cite{matsumoto2015estimate}, dialogue acts \cite{boyer2011affect}, and negotiation dialogue dynamics \cite{ghanem2021fakeflow}. 
In these papers, affective phenomena were 
also used to predict cognitive states such as personality traits \cite{10.1145/2388676.2388689, sogancioglu2021can}, working memory \cite{gabana2017effects}, cognitive task performance (especially tasks that require high cognitive-overload) \cite{kalatzis2022emotions}, and perceptions of other individuals  \cite{mcduff2013measuring, siddiquie2015exploiting}. We find that the main application domains were \textbf{healthcare} and \textbf{education}. The papers with healthcare applications used affective phenomena to predict depression severity \cite{scherer2013audiovisual}, suicidal ideation \cite{sawhney2020time, sawhney2021phase, sawhney2021suicide}), schizophrenic behavior \cite{hong2012lexical}, patient satisfaction with doctor communication styles \cite{sen2017modeling}, and multimodal distress assessment in patients during dyadic patient-clinician interactions, where affective context of clinician questions was taken into account \cite{ghosh2014multimodal}. The papers with education applications used affective phenomena to predict children's pronunciation ability in an educational reading context \cite{8925515} and cognitive strategies during learning \cite{cloude2021negative}). Additional papers focused on using affective features to predict presentation proficiency \cite{ramanarayanan2015evaluating}, speaker reliability \cite{parthasarathy2017predicting}, and other affective states \cite{10.1145/3536221.3556584}.

\textbf{Affective Phenomena in an Auxiliary Task:} 13 out of the 52 papers used affective phenomena in auxiliary modeling tasks (e.g., emotion prediction) to improve performance in downstream tasks predicting human social behaviors and cognitive states. 10 of these 13 papers were from NLP venues. Affective phenomena were used in auxiliary tasks during pretraining, multi-task learning, and fine-tuning; models with these auxiliary tasks achieved improved performance in predicting the following states: stress \cite{turcan2021emotion, yao2021muser},  dialogue acts,\cite{saha2020towards, saha2021towards, sawhney2021multitask}, abusive behavior (e.g., harassment) \cite{rajamanickam2020joint}, stance \cite{zhang2020enhancing}, sarcasm, \cite{chauhan2020sentiment}, metaphor expression \cite{dankers2019modelling}, group cohesion \cite{maman2021using}, rhetorical behavior (e.g., critical, discriminative, supportive rhetoric) \cite{huguet2021us}, 
formality \cite{chawla2019pre}, frustration \cite{chawla2019pre}, and politeness \cite{chawla2019pre}. We also found one paper that used multi-task learning for both emotion shift prediction and dialogue act recognition, to improve the emotion recognition in multi-party conversations \cite{10.1145/3536221.3556601}.

\textbf{Affective Phenomena as Latent States:} 10 of these 52 papers treated affective phenomena as a latent state in models of human social behaviors and cognitive states. Several papers modeled group interaction dynamics by treating affective phenomena as latent states. Affective information was used as a latent variable in a partially observable Markov decision process to model dyadic human interactions, with  applications in tutoring systems \cite{hoey2013bayesian}. Other papers defined interpersonal emotion networks as graphs capturing relationship dynamics in multi-party settings \cite{10.1145/1647314.1647333} and modeled emotion under the premise of emotion states modulating dyadic human behavior \cite{yang2017weighted}. Additional papers viewed affective phenomena as latent states in models for personality \cite{wache2015implicit},  moral conflicts \cite{lee2017exploring}, creative performance \cite{morris2013affect}, human sarcasm perception \cite{oprea2022should}, toxicity perception \cite{lahnala2022mitigating}, and decision-making \cite{barthet2022play}, as well as engagement, interactivity, impatience, reﬂectivity, and cognitive learning outcomes during online education \cite{afzal2017abc}. 

The 52 papers are listed in the Appendix \textbf{Table \ref{func_papers}}.

\section{Summary and Future Directions} 
\label{future_recommendations}
In this paper, we explore \textit{how}, and \textit{to what extent}, research at the intersection of multimodal interaction and affective computing has  treated the role of affective phenomena in AI systems. In our  sample of over 16,000 papers from ICMI, ACII, ACL, and EMNLP, we identify 910 papers related to affective phenomena (affect and emotion) and find that this body of research  has primarily focused on enabling machines to recognize and express
affect and emotion. We find that the use of affect and emotion to enhance machine understanding of human social behaviors and cognitive
states has been understudied (52 of the 910 papers); this role is visualized in \textbf{Figure \ref{fig:teaser}}. However, we observe an  emerging interest in this direction (\textbf{Figure \ref{fig:cumulative_func}}). From our analysis, we offer insights into future directions to expand the role of affective phenomena in AI systems. 

\textbf{(1) Expanding the Roles of Affective Phenomena in Multimodal Interaction Models:} We identify an emerging area of research using affective phenomena as features, in auxiliary tasks, and as latent states to improve downstream models of social behaviors and cognitive states. We acknowledge that there may be a selection bias influencing the publication of papers with positive results. We recommend that future research efforts replicate and validate past empirical findings on the usefulness of affective phenomena in these modeling contexts. From our study, we found that 10 of the 13 papers that used affective phenomena in auxiliary tasks were from experiments in unimodal text-only settings. We recommend that future research efforts explore affective signals in auxiliary tasks to include affective context in \textit{multimodal} models of human social behaviors and cognitive states during different stages of the modeling process (e.g., pretraining, fine-tuning, co-learning \cite{zadeh2020foundations}). In addition, we suggest that future multimodal interaction research explore the inclusion of explicit and implicit affective signals as rewards under the reinforcement learning from human feedback framework \cite{9131765, christiano2017deep}), to adapt and expand AI system understanding of human social behaviors and cognitive states.

\textbf{(2) Cognitively and Neurally-Inspired Models that Reflect the Complex Interaction of Affect, Behavior, and Cognition:} The use of affective phenomena in AI models of human social behaviors and cognitive states can be motivated by existing relationships among these three constructs in humans \cite{adolphs2001interaction, ito2001affect, bechara1994insensitivity, niedenthal2000emotional, dukes2021rise} (overview in \textbf{Section \ref{abc}}). The existence of these theorized and empirically-validated relationships in psychology and neuroscience (e.g., affect-as-information, affect priming, and affect infusion theories \cite{clore2001affect, clore2013affective, klauer2003affective, forgas2013affect}) have the potential to inform the science of multimodal interaction research. We, therefore, recommend that future research efforts explore how to leverage these theories to build computational models that reflect the complex interaction among these three constructs in humans. We believe that cognitively and neurally-inspired models have the potential to advance the ability of AI systems to use affective phenomena in order to better understand human social behaviors and cognitive states. 

\textbf{(3) Expanding Multimodal Social Interaction Contexts:}
We find that a majority of the papers in our sample focus on monadic contexts, with a limited focus on challenges present in dyadic and multi-party contexts. We recommend that future research efforts explore how AI systems can better integrate affective state predictions from multiple people and multiple contexts to inform models of social behaviors and cognitive states in group-level dynamics. For example, how can an AI system integrate its affect predictions of both people in a dyad in order to predict their synchrony (group-level behavior) \cite{wood2021forms}? How can affective state predictions of one person in a multi-party interaction inform an AI system's understanding of behavior and cognition in other participants?

\textbf{(4) Expanding Application Areas: }
We find that several of the papers successfully used affective phenomena to predict social behaviors and cognitive states in the domains of \textit{healthcare} (e.g., modeling depression, doctor-patient communication dynamics) \cite{scherer2013audiovisual, sawhney2020time, sawhney2021suicide, sawhney2021phase, sen2017modeling, ghosh2014multimodal, turcan2021emotion, yao2021muser} and \textit{education} (e.g., modeling student learning dynamics) \cite{8925515, cloude2021negative, afzal2017abc}. We recommend that future research efforts further explore how affective phenomena can be used as features, auxiliary tasks, and latent states to improve AI systems that support human health, education, and well-being through applications and empirical validation in additional populations and social contexts. Since it is possible that affective phenomena might not be useful to inform models in all settings, we recommend that future research explore techniques to enable AI systems to efficiently estimate when they need to \textit{query} affective state information from their environment in order to improve their understanding of the social behaviors and cognitive states of the people around them.

\begin{acks}
This material is based upon work supported by the National Science Foundation (NSF) Graduate Research Fellowship Program under Grant No. DGE2140739. Any opinions, findings, and conclusions or recommendations expressed in this material are those of the authors and do not necessarily reflect the views of the NSF.
\end{acks}

\bibliographystyle{ACM-Reference-Format}
\bibliography{sample-base}

\newpage

\appendix
\section{Appendix}
The 52 papers that use affect or emotion to enhance machine understanding of social behaviors and cognitive states are listed on the next page in \textbf{Table \ref{func_papers}}.

\onecolumn
    \begin{longtable}
    {l|l|p{5in}|l}
    \caption{The 52 papers that use affect or emotion to enhance machine understanding of social behaviors and cognitive states.\label{func_papers}}\\
        \textbf{Year} & \textbf{Venue} & \textbf{Title} & \textbf{Use} \\
        \hline
        \endfirsthead
         \caption{(continued) The 52 papers that use affect or emotion to enhance machine understanding of social behaviors and cognitive states.\label{func_papers}}\\
        \textbf{Year} & \textbf{Venue} & \textbf{Title} & \textbf{Use} \\
        \hline
        \endhead
        2009 & ACII &  Learning models of speaker head nods with affective information \cite{lee2009learning} & Feature \\ 
        2009 &ICMI &  Recognizing communicative facial expressions for discovering interpersonal emotions in group meetings \cite{10.1145/1647314.1647333}  & Latent State \\ 
        2011 & ACL & An Affect-Enriched Dialogue Act Classification Model for Task-Oriented Dialogue \cite{boyer2011affect} & Feature \\
        2012 &ICMI &  FaceTube: predicting personality from facial expressions of emotion in online conversational video \cite{10.1145/2388676.2388689} & Feature \\ 
        2012 & ACL & Self-Disclosure and Relationship Strength in {T}witter Conversations \cite{bak2012self} & Feature \\
        2012 & EMNLP & Lexical Differences in Autobiographical Narratives from Schizophrenic Patients and Healthy Controls \cite{hong2012lexical} & Feature \\
        2013 &ACII &  Affect and Creative Performance on Crowdsourcing Platforms \cite{morris2013affect}& Latent State \\ 
        2013 &ACII &  Bayesian Affect Control Theory \cite{hoey2013bayesian} & Latent State \\ 
        2013 &ACII &  Measuring Voter's Candidate Preference Based on Affective Responses to Election Debates \cite{mcduff2013measuring} & Feature \\ 
        2013 &ICMI &  Audiovisual behavior descriptors for depression assessment \cite{scherer2013audiovisual}  & Feature \\ 
        2014 &ICMI &  A Multimodal Context-based Approach for Distress Assessment \cite{ghosh2014multimodal} & Feature \\ 
        2014 & EMNLP & Classifying Idiomatic and Literal Expressions Using Topic Models and Intensity of Emotions \cite{peng2018classifying} & Feature \\
        2015 &ACII &  Emotional signaling in a social dilemma: An automatic analysis  \cite{stratou2015emotional}& Feature\\ 
       2015 & ACII &  Estimate the intimacy of the characters based on their emotional states for application to non-task dialogue \cite{matsumoto2015estimate} & Feature \\ 
        2015 &ICMI &  Evaluating Speech, Face, Emotion and Body Movement Time-series Features for Automated Multimodal Presentation Scoring \cite{ramanarayanan2015evaluating} & Feature \\ 
       2015 & ICMI &  Exploiting Multimodal Affect and Semantics to Identify Politically Persuasive Web Videos \cite{siddiquie2015exploiting}  & Feature \\ 
        2015 & ICMI &  Implicit User-centric Personality Recognition Based on Physiological Responses to Emotional Videos \cite{wache2015implicit} & Latent State \\ 
        2016 & ACL & Metaphor Detection with Topic Transition, Emotion and Cognition in Context \cite{jang2016metaphor} & Feature \\
        2017 & ACII & Effects of valence and arousal on working memory performance in virtual reality gaming \cite{gabana2017effects}  & Feature \\ 
       2017 & ACII &  Modeling doctor-patient communication with affective text analysis \cite{sen2017modeling}  & Feature \\ 
        2017 & ACII &  Weighted geodesic flow kernel for interpersonal mutual influence modeling and emotion recognition in dyadic interactions \cite{yang2017weighted}  & Latent State \\ 
        2017 & ACII &  The ABC of MOOCs: Affect and its inter-play with behavior and cognition \cite{afzal2017abc} & Latent State  \\ 
        2017 & ACII &  Exploring moral conflicts in speech: Multidisciplinary analysis of affect and stress \cite{lee2017exploring} & Latent State \\ 
        2017 & ACII &  Predicting speaker recognition reliability by considering emotional content  \cite{parthasarathy2017predicting} & Feature \\ 
        2019 & ACII &  Pre-trained Affective Word Representations \cite{chawla2019pre} & Auxiliary Task\\ 
        2019 & ACII &  Frustratingly Easy Personalization for Real-time Affect Interpretation of Facial Expression \cite{8925515} & Feature \\ 
        2019 & EMNLP & Modelling the interplay of metaphor and emotion through multitask learningn \cite{dankers2019modelling} & Auxiliary Task  \\
        2020 & EMNLP &  A Time-Aware Transformer Based Model for Suicide Ideation Detection on Social Media \cite{sawhney2020time} & Feature \\ 
         2020 & ACL &  Towards Emotion-aided Multi-modal Dialogue Act Classification \cite{saha2020towards} & Auxiliary Task \\ 
        2020 & ACL &  Joint Modelling of Emotion and Abusive Language Detection \cite{rajamanickam2020joint}  & Auxiliary Task \\ 
        2020 &ACL &  Enhancing Cross-target Stance Detection with Transferable Semantic-Emotion Knowledge  \cite{zhang2020enhancing} & Auxiliary Task \\ 
        2020 & ACL & Sentiment and Emotion help Sarcasm? A Multi-task Learning Framework for Multi-Modal Sarcasm, Sentiment and Emotion Analysis \cite{chauhan2020sentiment}  & Auxiliary Task \\ 
        2020 & ICMI & Introducing Representations of Facial Affect in Automated Multimodal Deception Detection \cite{mathur2020introducing}  & Feature \\ 
        2021 & ACII &  Towards Emotion-Aware Agents For Negotiation Dialogues \cite{chawla2021towards}  & Feature \\ 
        2021 & ACII &  Using Valence Emotion to Predict Group Cohesion’s Dynamics: Top-down and Bottom-up Approaches \cite{maman2021using} & Auxiliary Task \\ 
       2021 &  ACII & Negative emotional dynamics shape cognition and performance with MetaTutor: Toward building affect-aware systems \cite{cloude2021negative}  & Feature \\ 
       2021 & ACII & Can mood primitives predict apparent personality? \cite{sogancioglu2021can} & Feature\\
        2021 &  EMNLP & CHoRaL: Collecting Humor Reaction Labels from Millions of Social Media Users \cite{yang2021choral}  & Feature \\ 
       2021 & ACL &  MUSER: MUltimodal Stress detection using Emotion Recognition as an Auxiliary Task \cite{yao2021muser}  & Auxiliary Task \\ 
        2021 & ACL &  Us vs. Them: A Dataset of Populist Attitudes, News Bias and Emotions \cite{huguet2021us}  & Auxiliary Task \\ 
        2021 & ACL & Towards Sentiment and Emotion aided Multi-modal Speech Act Classification in Twitter \cite{saha2021towards}  & Auxiliary Task\\ 
        2021 & ACL & Multitask Learning for Emotionally Analyzing Sexual Abuse Disclosures \cite{sawhney2021multitask} & Auxiliary Task \\ 
        2021 & ACL & Emotion-Infused Models for Explainable Psychological Stress Detection \cite{turcan2021emotion}  & Auxiliary Task \\ 
       2021 &  ACL & PHASE: Learning Emotional Phase-aware Representations for Suicide Ideation Detection on Social Media \cite{sawhney2021phase} & Feature \\ 
        2021 & ACL & Fake Flow: Fake News Detection by Modeling the Flow of Affective Information \cite{ghanem2021fakeflow} & Feature \\ 
        2021 & ACL & Suicide Ideation Detection via Social and Temporal User Representations using Hyperbolic Learning \cite{sawhney2021suicide} & Feature \\
        2022 & ACII & Play with Emotion: Affect-Driven Reinforcement Learning  \cite{barthet2022play}& Latent State\\ 
        2022 & ACL & Should a Chatbot be Sarcastic? Understanding User Preferences Towards Sarcasm Generation \cite{oprea2022should}  & Latent State \\ 
       2022 & ACL &  Mitigating Toxic Degeneration with Empathetic Data: Exploring the Relationship Between Toxicity and Empathy \cite{lahnala2022mitigating} & Latent State \\ 
        2022 &ICMI &  Emotions Matter: Towards Personalizing Human-System Interactions Using a Two-layer Multimodal Approach \cite{kalatzis2022emotions}  & Feature \\ 
        2022 & ICMI & Real-Time Multimodal Emotion Recognition in Conversation for Multi-Party Interactions \cite{10.1145/3536221.3556601} & Auxiliary Task \\ 
        2022 &ICMI &  Supervised Contrastive Learning for Affect Modelling  \cite{10.1145/3536221.3556584} & Feature \\ 
    \end{longtable}

\end{document}